\documentclass[pra,showpacs,onecolumn,byrevtex,superscriptaddress]{revtex4}
\usepackage{graphicx}
\usepackage{subfigure}
\usepackage{dcolumn}
\usepackage{amsmath}
\usepackage{epsfig}
\usepackage{color}
\usepackage{amsmath}
\usepackage{multirow}
\usepackage{caption}

\begin{document}
\title{ Tagging idea in  continuous variable quantum key distribution and its application }

\author{Chenyang Li}
\email{chenyangli@ece.utoronto.ca}
\affiliation{ Department of Electrical \& Computer Engineering,University of Toronto, Toronto, M5S 3G4, Canada}

\author{Thomas Van Himbeeck }
\affiliation{ Department of Electrical \& Computer Engineering,University of Toronto, Toronto, M5S 3G4, Canada}

\author{Li Qian}
\affiliation{ Department of Electrical \& Computer Engineering,University of Toronto, Toronto, M5S 3G4, Canada}

\author{Hoi-Kwong Lo}
\affiliation{ Department of Electrical \& Computer Engineering,University of Toronto, Toronto, M5S 3G4, Canada}
\affiliation{Center for Quantum Information and Quantum Control, Department of Physics, University of Toronto, Toronto, M5S 3G4, Canada}

\begin{abstract}
Despite tremendous theoretical and experimental progress in continuous variable (CV) quantum key distribution (QKD), its security has not been rigorously established for  practical systems with multiple imperfections. The idea of   tagging is widely used in  security proofs of discrete variable quantum key distribution with imperfect devices. In this paper, we generalize the tagging idea from discrete variable to continuous variable.
 Based on untagged signals,  we  prove the security of the imperfect quantum key distribution system in the most conservative case. By introducing a generic imperfection model, we can evaluate and further incorporate multiple imperfections in the different stages such as modulation, channel and detection. Finally, with this generic model and tagging idea, we can prove the security of continuous variable key distribution system with multiple imperfections.
Our case study shows our proofs are able to provide secure keys in the presence of both modulation and detection flaws.
\end{abstract}

\maketitle

\section{Introduction}
\normalsize

Quantum key distribution(QKD) allows two distant parties, Alice and Bob, to share a common string of secret data \cite{Lo2014,Weedbrook2012,Diamanti2015}. Based on the laws of quantum mechanics, QKD offers information-theoretic security.  Generally speaking, key distribution protocols  can be mainly divided into two groups: discrete variable quantum key distribution(DV QKD)\cite{Lo2005,Ma2005} or continuous variable quantum key distribution(CV QKD)\cite{Jouguet2013,Lodewyck2005,Qi2007,Liu2017}. In DV QKD systems, the key information is encoded on  the properties of single photons\cite{Lo2014}, e.g. polarization. In CV QKD systems, the key information is carried by  the properties of light that are continuous such as the value of the quadrature of a coherent state \cite{Diamanti2015,Weedbrook2012}. CVQKD has the potential for high key rate and low cost implementations using standard telecom components such as homodyne detectors\cite{Diamanti2015}.  Recently, a silicon photonic chip platform for CV QKD has demonstrated a stable, miniaturized and low-cost system  that is compatible with the existing fibre
optical communication infrastructure\cite{Liu2019}.  On the other hand, the fiber-based CVQKD experiment has demonstrated the secret key transmission over more than 200 km\cite{Hong2020}.

Despite the enormous progress in the field of QKD,  the most important question in quantum communication is always how secure CV QKD really is.  Unfortunately, the security research of CV QKD has fallen behind that of DV QKD. Just like DV QKD, CV QKD is vulnerable to security loopholes due to imperfect devices. The deviation from the idealized assumptions in the state preparation\cite{Liu2017}, e.g. imperfect Gaussian variable modulation, may leak some information to an eavesdropper. Another fact in CV QKD proofs  is that we rarely take attacks on the phase reference signal, local oscillator(LO), into account\cite{Diamanti2015}.  Imperfections on the detection stage also open the door to potential security loopholes.
Unlike DV QKD where a lot of research efforts have been spent\cite{Lo2005,Lo2016,Tamaki2014}, these imperfection in CV QKD remain to be explored in depth\cite{Weedbrook2012,Diamanti2015}, especially if multiple imperfections exist.

In this work, we will  prove the security of CV QKD  with multiple imperfections by the  tagging idea. First, we will develop the  theory of tagging idea in the CVQKD , which is originally proposed to prove the security of DV QKD with imperfect devices\cite{Gottesman2004}, and  recently applied to the CVQKD  with intensity fluctuating sources\cite{Li2020}. Here, we will clarify the similarities and differences when we apply the tagging idea in the CV QKD and DV QKD. Similarly in both CV and DV, by dividing up signals into  untagged and tagged signals , we can obtain the secret key from untagged signals after error correction and privacy amplification process.  Different from DV, the untagged states in CV will be defined based on a prescribed region which is chosen by Alice and Bob. In other words, Alice and Bob will choose a conservatively secure region and tagged states will  possibly still hold some secrecy. Even though there could be a small sacrifice in the secret key, the tagging idea will  significantly simplify the proof for CVQKD with imperfect devices.  Next, by introducing a generic imperfection model, we can prove the security of a CV QKD system with multiple imperfections existing simultaneously.
Our proofs  are simple to implement without any hardware adjustment for  continuous variable quantum key distribution system. In the end, we demonstrate a case study by using our method to analyze the security of  a CV QKD system with two types of imperfections in the modulation and detection stages.

\section{Tagging idea in discrete variable quantum key distribution}
Here, we recall the tagging idea in discrete variable quantum key distribution\cite{Gottesman2004}. A source with flaws may "tag" some of the qubits with information which reveals the  basis  used in the preparation to the eavesdropper. By using this basis information, an eavesdropper can collect encoded bit information without causing any disturbance.  In other words, Alice and Bob should eliminate all information for the tagged state in the privacy amplification process.
A typical case to use tagging idea is a source emitting weak coherent states. With noneligible probability, the source will emit multiple photons. By photon number splitting attack, eavesdropper can split the extra photons to collect the bit information without causing any extra errors, which will compromise the security of quantum key distribution if the multiple photons are not carefully taken into account.

Let us take the single photon source as a starting point. Suppose  that a fraction $p$ of  the qubits are tagged and  the total key length generated after error correction is $s$.
 The key length generated from tagged and untagged states are, respectively, $p s$ and $(1-p) s$. Next, we can imagine executing privacy amplification on two different strings. Since the privacy amplification scheme for quantum key distribution is linear, such as applying one parity check matrix to the sifted key after error correction, the key obtained is the bitwise XOR
\begin{gather}\label{1}
  s_{untagged}\bigoplus s_{tagged}.
\end{gather}
If $s_{untagged}$ is private and random, the sum will always be private and random, even if eavesdropper knows everything about $s_{tagged}$.
Therefore, after privacy amplification, the final secret key length will be
\begin{gather}\label{1}
  K=s_{untagged}(1-H_2(\delta))=(1-p)s(1-H_2(\delta))
\end{gather}
where $\delta$  is the phase error rates for untagged states. Only key length ,$(1-p)s$, generated from untagged states should be maintained, and phase errors should be corrected.

Next, let us consider a source emitting weak coherent states.  Single photon and multiple photons are, respectively,  the untagged state and tagged states. Since the single photon is indistinguishable from multi photons by Alice and Bob, Alice and Bob can not choose any post-selection scheme. However, based on the probability of single photons, Alice and Bob can estimate the gain of single photons.
Therefore, after privacy amplification, the final secret key length will be
\begin{gather}\label{1}
  K=Q_1\{1-H_2(e_{pahse})\}-f_uQ(u)E(u)
\end{gather}
where $Q_1$ is the gain of single photon, $e_{pahse}$ is the phase error rate of single photon, $f_u$ is the error correction coefficient, $Q(u)E(u)$ is the bit errors for all states.

From this equation, we can see that it is not necessary for Alice and Bob to physically distinguish the single photon and multiple photons once the probability of single photon can be estimated. Only key generated from single photon will be kept and the phase error of single photon will be corrected. On the other hand,  due to the indistinguishability from single photon to multiple, all bit errors should be corrected.

\section{Tagging idea in continuous variable quantum key distribution}
Here, we will apply the tagging idea to continuous variable quantum key distribution.    The motivation is that the tagging idea is useful to help prove the security of quantum key distribution with  imperfect devices assumptions.  In CVQKD, we define  the untagged  states as the states from which Alice and Bob will not overestimate cost of privacy amplification by their data, which is determined by the number of phase errors in the DVQKD.  In other words, we will assume the worst case and the untagged states are always conservative secure after privacy amplification. Just like the DVQKD, we will analyze the security following the process of error correction and privacy amplification.

Suppose  that a fraction $p$ of  the states are tagged. If the reverse reconciliation is considered,
the total entropy used to generate the key is $H(X_B)$. The error correction cost is $H(X_B|X_A)$.  The raw key rate generated from  untagged states after (bit) error correction is
\begin{gather}\label{1}
  R=p H(X_B)- H(X_B|X_A).
\end{gather}
Since the tagged state and untagged states are indistinguishable with each other, Alice and Bob should perform error correction for both states. This is expected as in DVQKD, Alice and Bob should correct the bit errors for both single photon (untagged state) and multi photons(untagged states).
Next, we consider the privacy amplification in CVQKD which corresponds to the phase error correction in DVQKD. In DVQKD, Alice and Bob only need to correct the phase errors in the single phonon. Similarity, in CVQKD, Alice and Bob only need to perform the privacy amplification in the untagged states, which result in the key rate as
\begin{gather}\label{1}
  R=p H(X_B)- H(X_B|X_A)-\chi_{BE, p}=I(AB)-(1-p) H(X_B)-\chi_{BE, p}.
\end{gather}
where $\chi_{BE, p}$ represents the total privacy amplification cost of the untagged states. Based on our definition that Alice and Bob will not overestimate the privacy amplification cost of the untagged  states by their data, which means that
\begin{gather}\label{1}
  \chi_{BE, p} \leq p \chi_{BE},
\end{gather}
where $\chi_{BE}$ is the cost of privacy amplification which is directly measured from the Alice's and Bob's  the encoded and measured data.
Finally, given the reconciliation efficiency $\beta$, the secret key rate can be shown as
\begin{gather}\label{1}
  R_r= \beta I(A:B)-(1-p) H(X_B)-p\chi_{BE}.
\end{gather}

\section{security analysis of continuous variable quantum key distribution with multiple imperfections}

\subsection{Generic imperfection model for CV QKD stage }
\begin{figure}[!htb]
\centering
\includegraphics [width=100mm,height=18mm]{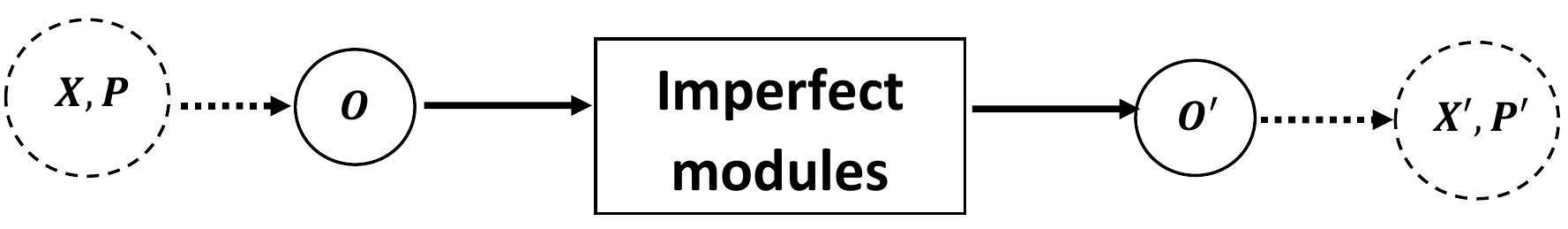}
\caption{Generic imperfection model }
\end{figure}\label{idac}
 Fig 1 shows the generic imperfection model for CV QKD stage. A signal state, $O$, carrying the encoding information, $X$ and $P$ , pass the imperfect modules and becomes a state, $O'$, carrying the encoding information ,$X'$ and $P'$.
A generic  model to describe the imperfection can be expressed as
\begin{gather}\label{1}
  \hat{X}\stackrel{f}{\longrightarrow}\hat{X}'
\end{gather}
where $f(\bullet)$  describes the quadrature transformation from $X$  to $X'$(or $P$ to $P'$).

Before we define the general transformation, let us review two transformation models for current CVQKD.

1) In a lossy channel, the input signals  are attenuated and combined with the thermal noise, and the quadrature transformation can be expressed as \cite{Weedbrook2012}
\begin{gather}\label{1}
  \hat{X}'=\sqrt{T}\hat{X}+\sqrt{1-T}\hat{X}_{th}
\end{gather}
where $\hat{X}_{th}$are the thermal state in the channel.

2) With phase rotation operator,  the quadratures are transformed via simple linear transformation \cite{Weedbrook2012}
\begin{gather}\label{1}
  \hat{X}'=\text{cos}\theta\hat{X}+\text{sin}\theta\hat{P}
\end{gather}
where $\theta $ is the angle of rotation.

Considering the fact that $\hat{X}_{th}$ and $\hat{P}$ are both independent on $\hat{X}$, these two transformations can be described in a generic way:
\begin{gather}\label{1}
  \hat{X}'=a\hat{X}+\sqrt{1-a}b
\end{gather}
where $a$ is a constant, $b$ is a random variable with variance.

Considering the fluctuation such as atmospheric effect of the channel and  phase fluctuation, the parameter $a$ can also become a random variable. Thus, we obtain a  generic model to describe the imperfect module:
\begin{gather}\label{1}
 \hat{X}'=a\hat{X}+\sqrt{1-a}b
\end{gather}
where $a$ and $b$ can, respectively,  be a random variable with some probability distributions.   By this model, we can describe lots of imperfections, such as intensity fluctuation of signal states\cite{Liu2017} or local oscillator \cite{Liang2013}, phase fluctuation\cite{Qi2007} and atmospheric effects\cite{zeng2018}.

\subsection{continuous variable quantum key distribution with multiple imperfections}

With single generic imperfection model, we can now establish process for CV QKD with multiple imperfections. Suppose  that the quadrature transformation in the modulation, channel and detection can, separately, be described as  $f_m(\bullet), f_c(\bullet),f_d(\bullet)$, where
\begin{gather}\label{1}
f_m(X)= a_mX+\sqrt{1-a_m} b_m \\
f_c(X)= a_cX+\sqrt{1-a_c}b_c    \\
f_d(X)= a_dX+\sqrt{1-a_d}b_d   \\
\end{gather}
where  $a_m, b_m, a_c, b_c, a_d,b_d$ are all random variables with some probability distributions,

 The overall transformation can be expressed as
\begin{gather}\label{1}
 \hat{X_i}\stackrel{f_m}{\longrightarrow}\stackrel{f_c}{\longrightarrow}\stackrel{f_d}{\longrightarrow}\hat{X_o}
\end{gather}

\subsection{Tagging idea for  multiple imperfections}

In the quadrature transformations, $b_m,b_c$, and $b_d$ respectively represent a noise term in each stage. Bob's detection result $X_o$ will always reflect the overall noise effect by its variance.  In other words, the effect of $b_m,b_c$, and $b_d$ will  be included in the parameter estimation of noise term.
On the other hand, $a_m,a_c$, and $a_d$ represents a transmittance term. While Alice and Bob estimate the transmittance, they will only estimate $\bar{a}_m,\bar{a}_c,$ and $\bar{a}_d$, the average value of $a_m,a_c,a_d$. In other words, there could be some states passing a less attenuated or more attenuated channel compared to the  $\bar{a}_m,\bar{a}_c$, and $\bar{a}_d$. Now the question becomes that how to prove the security if we can only obtain the parameters of $\bar{a}_m,\bar{a}_c,$ and $\bar{a}_d$ while we cannot monitor individual state?
 The answer is that we can apply tagging idea from \cite{Gottesman2004} to prove the security of CV QKD.  We define  untagged Gaussian states that pass through the module with smaller amplitude $a$ than the mean value of each module. In other words, untagged Gaussian states will experience a high attenuation when compared to tagged Gaussian states. Therefore, the secret key rate generated from untagged Gaussian states will always be conservatively underestimated.  Note that we still need to correct the error for tagged Gaussian state.

In the transformation $f_m$, the state passing   a   process with parameter $a_m<\bar{a}_m$ is defined as the untagged Gaussian state, and then go into the transformation $f_c$;
In the transformation $f_c$, the state passing  a  process with parameter $a_c<\bar{a}_c$ is defined as the untagged Gaussian state, and then go into the transformation $f_d$;
In the transformation $f_d$, the state passing  a  process with parameter $a_d<\bar{a}_d$ is defined as the untagged Gaussian state; where $\bar{a}_m, \bar{a}_c,\bar{a}_d$ are respectively the mean value of each parameter.
Overall, the states passing a process with parameter $a_m<\bar{a}_m,a_c<\bar{a}_c, a_d<\bar{a}_d $ can be defined as  untagged Gaussian states.
The key question is how to choose the cut-off value for defining untagged signals. If the cut-off value is chosen to be the mean value of a Guassian distribution, then
in each stage of transformation, the probability of the untagged Gaussian state is 1/2 and the key rate is zero. So, we use a higher probability than 1/2 for untagged states by increasing the cut off value. If the cutoff is increased from $\bar{a}_m$ to $k\bar{a}_m$, Alice needs to map her data from $X_i$ to $kX_i$. Originally, this method is only proposed in the modulation stage\cite{Li2020}. Here, we will also  apply this method to the channel and detection stage.
If the cutoffs $a_m, a_c$, and $a_d$ are increased to $k_1a_m, k_2a_c,$ and  $k_3a_d$, Alice needs to map here data from  $X_i$ to $k_1k_2k_3X_i$.

Now, The asymptotic secret key rate extracted from untagged Gaussian states  for reverse
reconciliation can be shown  as
\begin{gather}\label{1}
  R_r= p(a_m < k_1\bar{a}_m)p(a_c< k_2\bar{a}_c)p(a_d< k_3\bar{a}_d)H(B)-H(B|A')-p(a_m < k_1\bar{a}_m)p(a_c<  k_2\bar{a}_c)p(a_d<  k_3\bar{a}_d)\chi_{BE}\\
     = I(A':B)-(1-p_0) H(X_B)-p_0\chi_{BE}.
\end{gather}
Here, $p_0$ is the probability of untagged Gaussian state and  $p_0= p(a_m <  k_1\bar{a}_m)p(a_c<  k_2\bar{a}_c)p(a_d<  k_3\bar{a}_d)$. $A'$ shows the Alice's recoded data after the mapping scheme, which help increase the probability of untagged states.
Given the reconciliation efficiency $\beta$, the secret key rate can be shown as:
\begin{gather}\label{1}
  R_r= \beta I(A':B)-(1-p_0) H(X_B)-p_0\chi_{BE}.
\end{gather}

\section{Application}
In this section, we will show the application of our security proof. In general, our security proof can deal with multiple imperfections simultaneously existing in the stages of modulation, channel and detection.  Without losing generality,  we will pick up the case which has two types of  imperfections: (1) the source has  intensity fluctuation (2)  the detector efficiency  has  fluctuations. We will assume Alice and Bob hold a Gaussian channel for key transmission.   The security analysis  can be divided into four steps.

Step 1: mathematically write down all  the transformations in the modulation, channel and detection. For example, in our case, the three transformations can be shown as
\begin{gather}\label{1}
  f_m(x)=a_mX, \\
  f_c(x)=a_cX+\sqrt{1-a_c}b_c, \\
  f_d(x)=a_dX+\sqrt{1-a_d}b_d
\end{gather}
For the transformation in the modulation, $a_m$  is a random variable and follows a distribution due to intensity fluctuation. For the transformation in the channel, $a_c$  is constant based on the channel transmittance, and $b_c$ is a random variable due to  channel excess noise.  For the transformation in the detection, $a_d$ is a random variable due to detector efficiency fluctuation, and $b_d$ is a random variable due to the electronic noise in the detector.
Based on our case assumptions, the two types of imperfection will only affect the parameters $a_m$ and  $a_d$.
  \begin{table}[!htb]
\centering
 \fontsize{12}{12}\selectfont
\caption{Evaluation parameters for CVQKD \cite{Lodewyck2007,Jouguet2014}}
\begin{tabular}{|c|c|c|c|c|}
\hline
 $\eta $      & $ \varepsilon_c $  & $v_{el}$ & $V_A $ & $\beta $  \\
\hline
0.60      &     0.02    & 0.02 &  18 & 95.6\\
\hline

\end{tabular}
\label{table1}
\end{table}

\begin{table}[!htb]
\centering
 \fontsize{12}{12}\selectfont
\caption{Parameters in the transformations}
\begin{tabular}{|c|c|c|c|c|c|}
\hline
   parameters      & $ a_m $  & $a_c$ &$a_d $&  $b_c $ & $b_d$  \\
\hline
   mean value      &    1    & $\sqrt{T_c}$ &  $\sqrt{\eta}$&  0 & 0\\
\hline
    variance      &    $V_1$    & 0&  $V_2$&  $\frac{T\varepsilon_c}{1-T_c}$ & $\frac{\eta v_{el}}{1-\eta}$ \\
\hline
\end{tabular}
\label{table1}
\end{table}

Step 2: obtain the mean value and variance of each parameters in the three transformations.
let us first list the parameters of a  CV QKD system for our evaluation in the table I,  where  $\eta$ is the mean value of detection efficiency,  $v_{el}$ is  the electronic noise of the homodyne detector, $\varepsilon_c$ is the excess noise in the channel, $V_A$ is the modulation variance and $\beta$ is the reverse reconciliation efficiency.
Given the channel transmittance $T_c$, now we can have a table II to show the mean values and variances of all parameters in the transformations, where $V_1$ and $V_2$ will be determined by our simulation model.

Step 3: determine your fluctuation model and optimize the probability of  the untagged states. For example, we choose the Gaussian distribution for the parameters $a_m$ and $a_d$.(Note that, you can also choose other distribution rather than Gaussian, such as uniform distribution \cite{Li2020}.) Also,  we choose the variance $V_1=0.0025$ and $V_2=0.0015$ for our simulation, which is around 5\% fluctuation. (Note that the mean value of the $a_c$ is different from $a_m$.) Next, optimize the cutoff coefficient, $k_1$ and $k_3$, of each distribution for the untagged states. (Note that this optimization process  should be combined with step 4.) The probability of untagged state can be shown as
\begin{gather}\label{1}
 p_0= p(a_m <  k_1\bar{a}_m)p(a_d<  k_3\bar{a}_d)=\int_{-\infty}^{k_1\bar{a}_m}\textrm{PDF}(a_m)da_m\int_{-\infty}^{k_3\bar{a}_d}\textrm{PDF}(a_d)da_d
\end{gather}
where $\bar{a}_m, \bar{a}_c,\bar{a}_d$ are respectively the mean value of each parameter.
With these optimal cutoff bounds, Alice should map her data from $X_i$ to $k_1k_3X_i$.

Step 4: compute the secret key rate.  After obtaining the probability of the untagged states,  we will use the Eq.(20) to compute the secret key rate.

\begin{figure}[!htb]
\centering
\includegraphics [width=100mm,height=80mm]{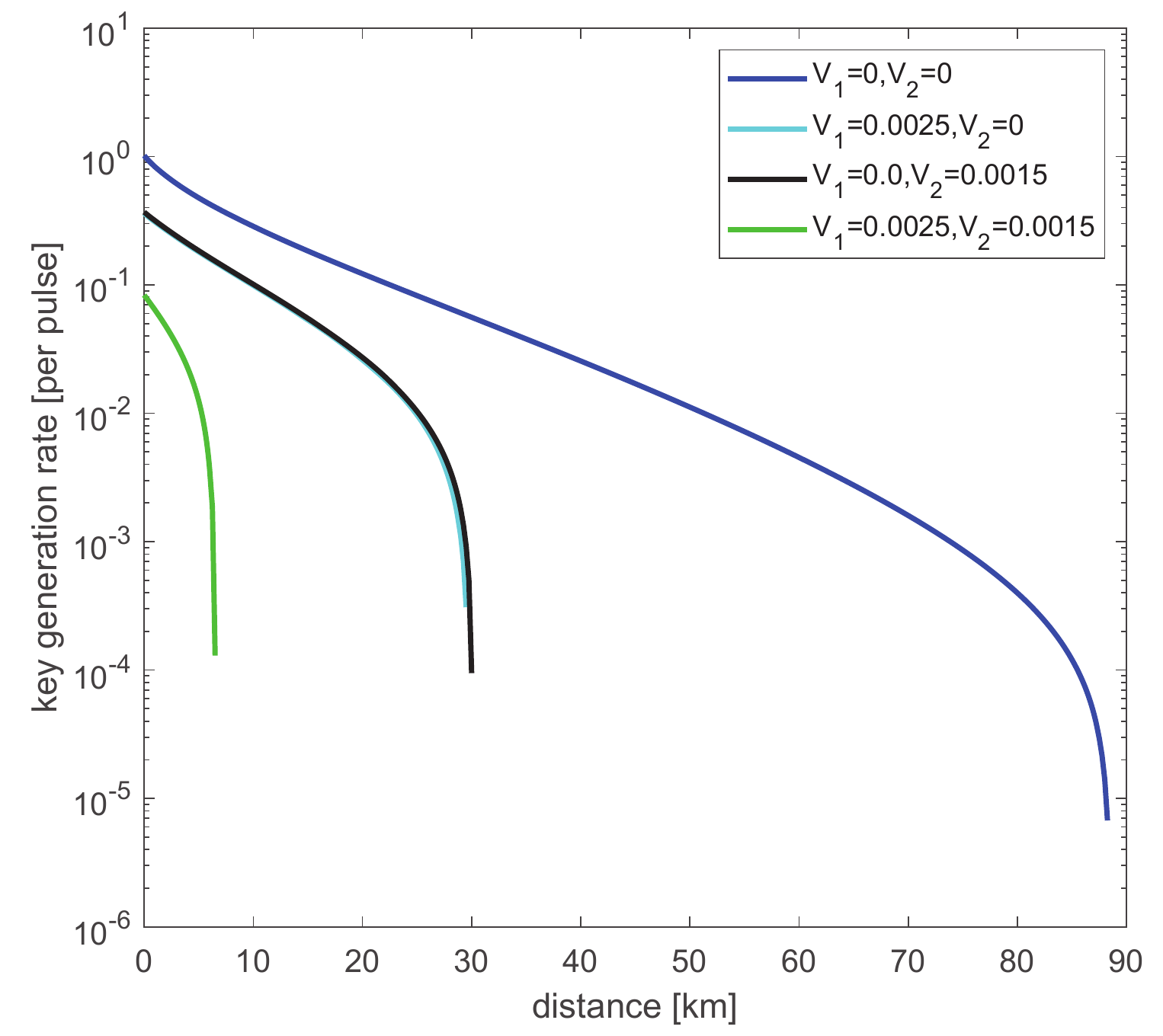}
\caption{Here, we compute the secret key rate versus distance. We compare four cases where the imperfections in the modulation and/or detection stage can/cannot exist. For our simulation, we set  both fluctuations   to be around 5\%. (Note that the variances are different due to the different mean value) With only one fluctuation, the maximum distance will drop from around 90 km to 30km. With both fluctuations, the maximum distance will drop to around 5 km.  }
\end{figure}\label{idac}

Fig 2 shows  the secret key rates over the distance. $V_1$ and $V_2$ is the variance in the table II. The fluctuation in the stages of  modulation and detection are set to be around 5\%. The parameters is listed in the table I.
It is expected that the key rate  drop fast when the system has both fluctuations. Another interesting observation is that the fluctuation in the modulation and detection stage will have a similar effect on the key rate, since the key rate is the same if there are only one type of fluctuation.

Fig 3 shows the secret key rates over the distance by using the parameters in \cite{Hong2020} for 202.81 km transmission. The parameters is shown in Table III.   With the two types of fluctuations in the stages of modulation and detection, the maximum transmission distance will be around 50km.
  \begin{table}[!htb]
\centering
 \fontsize{12}{12}\selectfont
\caption{Evaluation parameters for CVQKD in \cite{Hong2020}}
\begin{tabular}{|c|c|c|c|c|}
\hline
 $\eta $      & $ \varepsilon_c $  & $v_{el}$ & $V_A $ & $\beta $  \\
\hline
0.6134      &     0.0081    & 0.1523 &  7.65 & 98\\
\hline

\end{tabular}
\label{table1}
\end{table}

\begin{figure}[!htb]
\centering
\includegraphics [width=100mm,height=80mm]{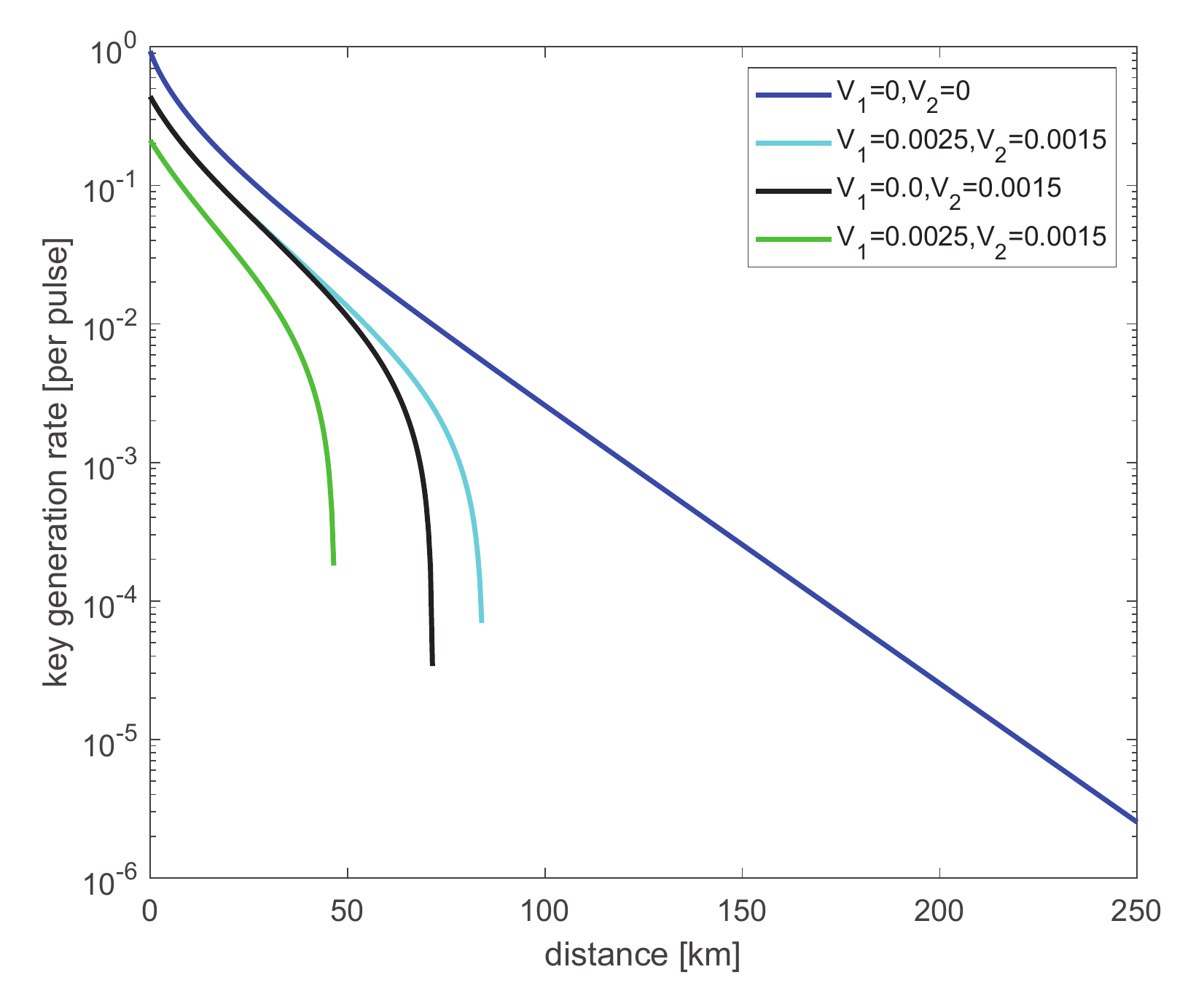}
\caption{Here, we compute the secret key rate versus distance by using the parameters in \cite{Hong2020}. }
\end{figure}\label{idac}

\section{Conclusion}

We have generalized the tagging idea in the CV QKD from the DV QKD.   We divide the signals into tagged and untagged signals, and the secret key will only be generated from untagged signals. After considering the  error correction cost and privacy amplification, the security of a CVQKD system can be proved. By proposing a generic imperfection model for CVQKD stage, we can describe different fluctuations simultaneous existing in a CV QKD system.  Based on the generic model and tagging idea, we prove the security of CVQKD with multiple imperfections. In the end, we use apply our security proof to a case which has two types of imperfections and obtain the secret key rates.

\section{ACKNOWLEDGEMENTS }\label{Sec7}
We acknowledge the financial support from the Natural Sciences and Engineering Research Council of Canada
(NSERC),  Huawei Technologies Canada Co., Ltd. Hoi-Kwong Lo acknowledges the financial support from the University of Hong Kong start-up fund.

\end{document}